# Ferromagnetic MnSn monolayer epitaxially grown on silicon substrate

Qian-Qian Yuan[1], Zhaopeng Guo[1], Zhi-Qiang Shi[1], Hui Zhao[1], Zhen-Yu Jia[1], Qianjin Wang[1], Jian Sun[1,2], Di Wu[1,2], Shao-Chun Li[1,2,3]*

[1] *National Laboratory of Solid State Microstructures, School of Physics, Nanjing University, Nanjing 210093, China*

[2] *Collaborative Innovation Center of Advanced Microstructures, Nanjing University, Nanjing 210093, China*

[3] *Jiangsu Provincial Key Laboratory for Nanotechnology, Nanjing University, Nanjing 210093, China*

* Corresponding authors: scli@nju.edu.cn (S.-C. Li)

ABSTRACT: Two-dimensional (2D) ferromagnetic materials have been exhibiting promising potential in applications, such as spintronics devices. To grow epitaxial magnetic films on silicon substrate, in the single-layer limit, is practically important but challenging. In this study, we realized the epitaxial growth of MnSn monolayer on Si(111) substrate, with an atomically thin Sn/Si(111)-$2\sqrt{3} \times 2\sqrt{3}$- buffer layer, and controlled the MnSn thickness with atomic-layer precision. We discovered the ferromagnetism in MnSn monolayer with the Curie temperature (Tc) of ~54 K. As the MnSn film is grown to 4 monolayers, Tc increases accordingly to ~235 K. The lattice of the epitaxial MnSn monolayer as well as the Sn/Si(111)-$2\sqrt{3} \times 2\sqrt{3}$ is perfectly compatible with silicon, and thus an sharp interface is formed between MnSn, Sn and Si. This system provides a new platform for exploring the 2D ferromagnetism, integrating magnetic monolayers into silicon-based technology, and engineering the spintronics heterostructures.

PACS: 75.70.Ak, 75.70.-i, 81.15. Hi, 68.37.Ef

Two-dimensional (2D) ferromagnetism demonstrated in the monolayer limit usually leads to the exotic properties in contrast to the bulk[1-5]. Not only that, interface engineering of the 2D magnetic films, for example coupling with semiconductors[6], topological insulator[7, 8], and many-body states[9, 10], etc, can bring versatile functionalities and thus be potential in technical applications[11-15]. The development towards integration and miniaturization pushed the magnetic films thinned down to the monolayer limit, as demanded by the motivation of low power and high speed devices[3-5]. Since the discovery of graphene in 2004, effort has long been devoted to induce magnetism in monolayered van der Waals (vdW) materials. It was initially attempted to extrinsically introduce local defects or proximity effect[16-20].

The intrinsic 2D ferromagnetism was firstly discovered in van der Waal monolayers of $CrI_3$[21] by Huang et al and $Cr_2Ge_2Te_6$[22] by Gong et al., respectively. In the following, more ferromagnetic monolayers were realized, such as the metallic $Fe_3GeTe_2$[2, 23] and especially the $MnSe_x$ monolayer with room temperature ferromagnetism[24]. The long-range magnetic order persists in the 2D limit presumably because the magnetic anisotropy overcomes the thermal fluctuations[25]. To date, most of the ferromagnetic vdW monolayers were achieved by mechanical exfoliation. Although the construction of lego-style vdW heterostructures may provide control of their properties[6, 26], to directly grow the epitaxial magnetic monolayer is practically important but rather challenging [24, 27-29], considering its high uniformity, tunability and the strategy to integrate with the mature silicon technology. From the view of fundamental science, epitaxial ferromagnetic monolayer provides an alternate opportunity to study the mechanism of 2D magnetism.

It has been predicted that the bulk MnSn in zinc-blende structure is a half-metallic ferromagnet, and if experimentally realized, MnSn is expected to be valuable in spintronics owing to its lattice compatibility with semiconductors[30, 31]. In this work, we succeeded in growing the monolayer MnSn on silicon substrate with a Sn/Si(111)-$2\sqrt{3} \times 2\sqrt{3}$ buffer layer by using molecular beam epitaxy (MBE). The lattices of MnSn and Si are compatible with a negligible mismatch, and thus a clean and sharp MnSn/Si interface is formed. The atomic structure of MnSn monolayer is determined by combination of scanning tunneling microscopy / spectroscopy (STM/STS) results and density functional theory (DFT) calculation, which is in fact a different structure from the bulk as previously reported. Magnetization measurement indicates that the monolayer MnSn undergoes a ferromagnetic transition with Tc of ~54 K. As the thickness is increased to ~4 monolayers, Tc increases up to ~235 K, close to room temperature. We would like to emphasize that even though the Sn buffer layer is introduced, its lattice is perfectly compatible with both Si and MnSn, and the Sn buffer layer is only two atomic-layer thin which is usually smaller than the typical length scale for spin injection / diffusion. It is thus the first-time realization of magnetic monolayer grown on silicon substrate with a sharp interface and atomic-layer controlled thickness.

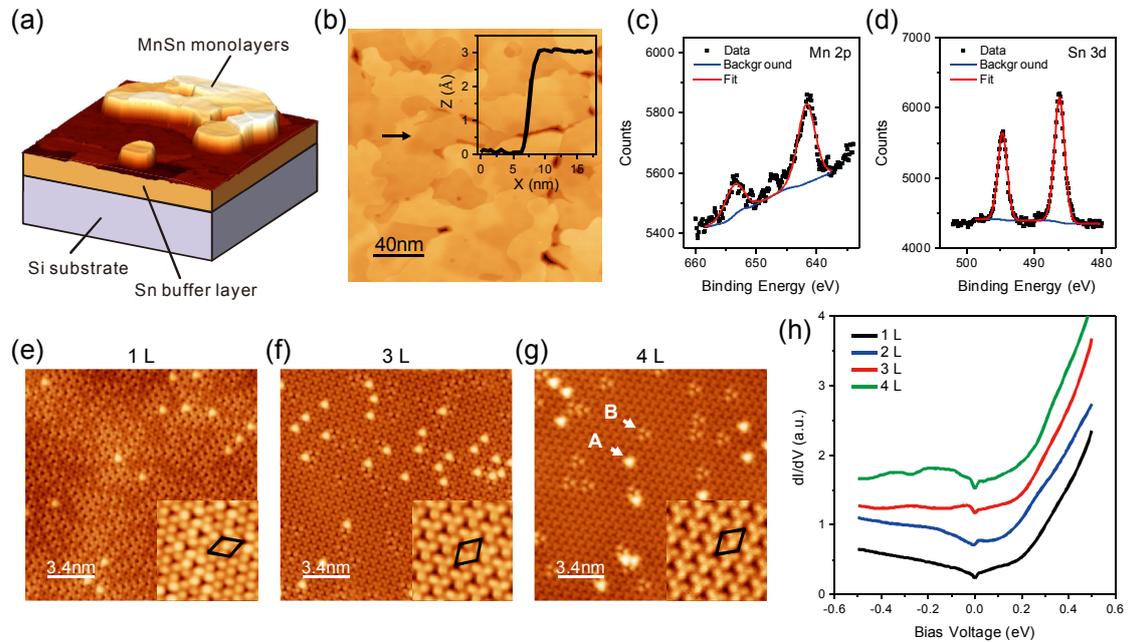

**Fig. 1.** Characterization of the epitaxial MnSn films on Sn/Si(111)-$2\sqrt{3} \times 2\sqrt{3}$. (a) 3D view of the STM topographic image (200 × 200 nm$^2$, $U$ = +1.5 V, $I_t$ = 50 pA) illustrating the growth of MnSn island on the Sn/Si(111)$2\sqrt{3} \times 2\sqrt{3}$ buffer layer. (b)

STM topographic image (200 × 200 nm$^2$) taken on the MnSn monolayer. $U$ = +1 V, $I_t$ = 100 pA. Inset: line-scan profile taken along the black arrowed line. The step height is measured to be ~3.0 Å. (c, d) XPS core level spectra of Mn 2p and Sn 3d. (e-g) High-resolution STM images taken on the 1L, 3L and 4L MnSn respectively. Insets: the corresponding atomic resolutions, black rhombuses mark the surface unit cells. The white arrows indicate two different kinds of defects, namely A and B. (h) Typical differential conductance (dI/dV) spectra measured on 1L to 4L at 4.2 K respectively.

The growth recipe of MnSn on Si substrate is as following: An atomically thin buffer layer of Sn/Si(111)-2√3×2√3 was firstly deposited on Si(111)-7×7 substrate, and then Mn and Sn were co-deposited on top of the Sn buffer layer. Figure 1(a) demonstrates a typical STM result of the MnSn islands obtained after depositing Mn and Sn at room temperature followed by annealing to ~500 °C. The Sn buffer layer still persists a 2√3×2√3 reconstruction after the MnSn growth, indicating that the Sn buffer layer works well to prevent intermixing with Si. By optimizing the annealing temperature, a high quality MnSn monolayer with large terraces was achieved, as shown in Fig. 1(b). The step height for the first monolayer is ~3.0 Å. The MnSn monolayer takes a hexagonal lattice symmetry, and the size of the unit cell is a = b = ~6.58 ± 0.07 Å, nearly half of that of the Sn/Si(111)-2√3×2√3 buffer layer (a = b = ~13.2 Å). As a result of the negligible lattice mismatch, a clear sharp interface is formed as observed in the HRTEM image (Fig. S1). Figure 1(e) shows the atomically resolved STM images taken on the MnSn monolayer. There exist three protrusions in each unit cell, ~0.34 nm apart from each other and forming a triangular morphology. Two kinds of defects, namely A and B, are identified, as marked by the white colored arrows in Fig. 1(g). The discussion below indicates that the defects A and B can be assigned as the adsorbed Sn atom and the interstitial Mn atom. Moreover, multilayered MnSn films can be also grown in an exact layer-by-layer fashion (Fig. S2). Atomically resolved STM morphology taken on the 3L and 4L films, as shown in Figs. 1(f) and 1(g), looks similar to that of monolayer with the same surface structure except for a reduced population of defects.

Differential conductance (dI/dV) curves, which is proportional to the local density of state (LDOS), were taken on the MnSn films as displayed in Fig. 1(h). More data can be found in Fig. S3. All the dI/dV curves taken across different layers show non-vanishing density of states in the bias range explored, indicating the metallic nature of MnSn films. However, there appears a tiny dip at the Fermi energy as temperature decreases from 78 K to 4.2 K. According to the STM topographic images, Figs. 1(e)-1(g) and Fig. S4, there is no apparent structural transition between 78 K and 4.2 K. Therefore, the dip possibly originates from the electron-electron interactions.

Before addressing the atomic structure of the MnSn monolayer, X-ray photoemission spectroscopy (XPS) characterization was firstly carried out to determine the chemical stoichiometry. Figures 1(c) and 1(d) present the Mn 2p and Sn 3d core-level spectra of the MnSn films of ~5 nm thick. The main peaks correspond to valence states in MnSn compound. Quantitative analysis indicates that the atomic ratio of Mn to Sn concentrations is ~1:1, consistent with the flux ratio for Mn to Sn.

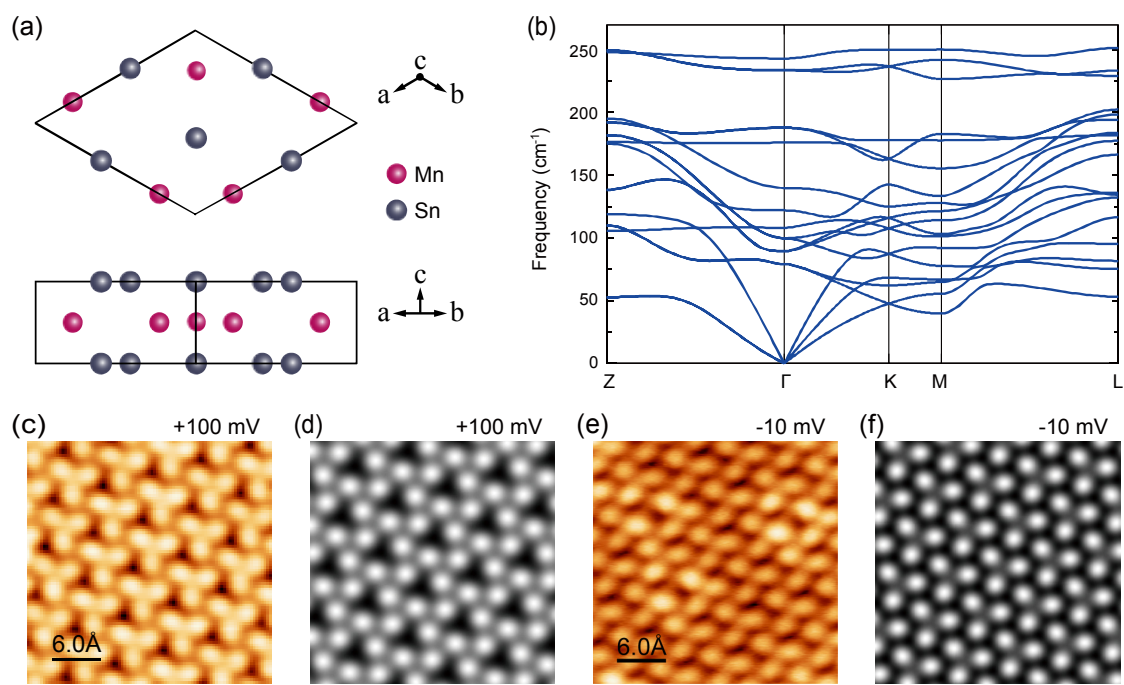

**Fig. 2.** Crystal structure model of the grown MnSn film. (a) Calculated atomic structure of the monolayer MnSn. Upper: top view, lower: side view. The red and gray balls represent respectively Mn and Sn atoms. (b) DFT calculated phonon spectra of MnSn, which shows the dynamical stability of the calculated structure model. (c, e) Atomic-resolution images taken at various bias voltages of $U$ = +100 mV and -10 mV. Bias voltages is indicated in the image, and the tunneling current is 100 pA. (d, f) DFT simulated images corresponding to the STM results (c, e).

We determine the atomic structure of the MnSn monolayer by aid of density functional theory (DFT). Based on the experimentally obtained lattice parameters and chemical stoichiometry, we performed a crystal structure searching for MnSn. The optimal crystal structure was choosen as illustrated in Fig. 2(a). The Mn and Sn atomic layers are stacked alternatively as Mn-Sn-Mn-Sn, and the Mn and Sn atoms form a distorted Kagome lattice with a hexagonal unit cell. The calculated lattice constants are a = b = 6.576 Å and c = 2.910 Å, in good agreement with the STM measurements. Details of the DFT optimized MnSn structure can be found in Supplementary Table S1. The Mn and Sn atoms in different layers are arranged to form triangles, respectively, especially for the Mn atoms, where the shortest Mn-Mn distance is only 2.572 Å. It is surprising that the shortest distance of Sn-Sn atoms (3.426 Å) in the same layer is larger than that of Sn-Sn atoms in different layer along the c direction (2.910 Å). This shows the structure has strong interlayer interaction, and thus has a high cleavage energy. However, it is still possible to stabilize a monolayer MnSn with substrate. The calculated phonon spectra of the MnSn phase shows no imaginary frequency (Fig. 2(b)), suggesting that the MnSn phase is dynamically stable.

To further confirm the validity of the calculated structure, we performed the DFT simulations of STM images. As shown in Figs. 2(c)-2(f), there exists a good agreement between the experimental and simulated images at various bias voltages. Such agreement can be further evidenced by the simulation of surface defects, as shown in supplementary information (Fig. S5). Based on the combination of DFT calculation and STM results, we conclude that the MnSn film grown on Sn/Si(111)-$2\sqrt{3} \times 2\sqrt{3}$ substrate holds the structure as shown in Fig. 2(a).

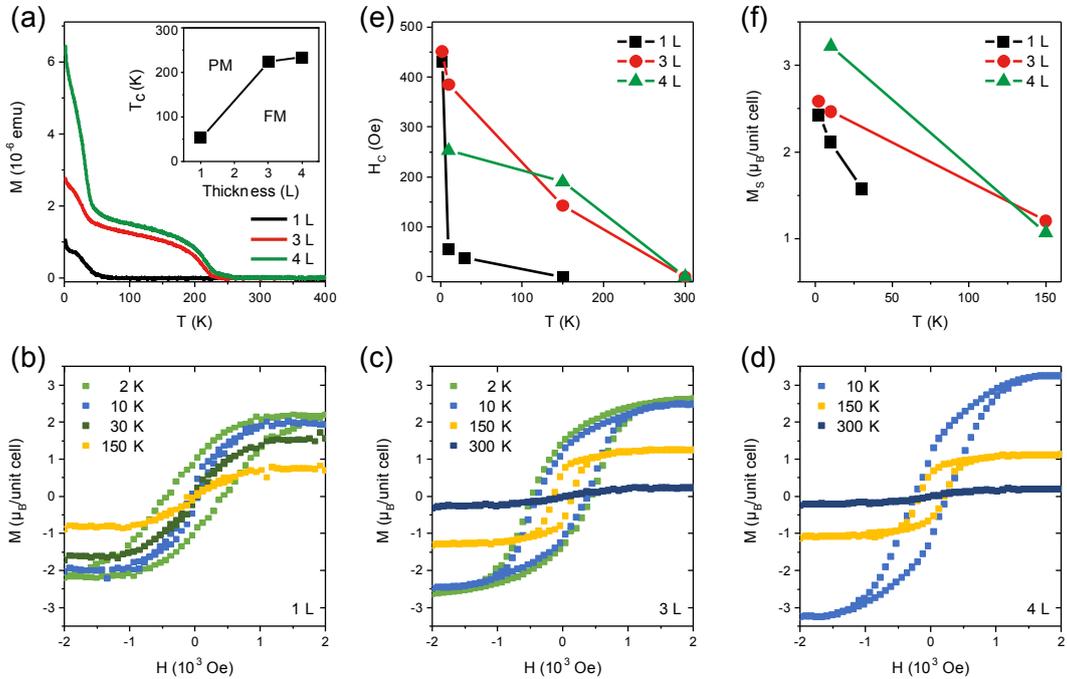

**Fig. 3.** Layer-dependent intrinsic ferromagnetism in MnSn. (a) Field cooling (FC) magnetizations as a function of temperature. The in-plane magnetic field (H) of 1000 Oe is applied. Inset: The corresponding Tc of 1L, 3L and 4L films derived from M~T

curves. (b-d) Magnetic hysteresis curves (M~H) obtained at various temperatures for different-thickness samples. The magnetization is normalized to magnetic moment per unit cell (each Mn atom). (e, f) Coercive force (Hc) and saturation magnetization (Ms) as a function of temperature. The surface morphologies and statistical distributions of the thickness are shown in Fig. S2. The coverages of the nominal 1L, 3L and 4L are determined to be 1.1 ML, 2.9 ML and 3.8 ML respectively.

Now we explore the magnetization properties of MnSn films. Field-cooling (FC) magnetization (M~T) measured on the nominal 1L, 3L and 4L films are plotted together in Fig. 3(a). During the measurement, a small parallel magnetic field of 0.1 T was applied to stabilize the magnetic moments. The background due to the diamagnetic response of Sn/Si(111)-$2\sqrt{3} \times 2\sqrt{3}$ substrate and Sn capping layers is subtracted (The raw data and background subtraction can be found in Figs. S6 and S7). The magnetic moment is positive at low temperature, and suppressed towards zero as temperature increases owing to the thermal fluctuations, indicating a ferromagnetic to paramagnetic transition. We defined the Curie temperature (Tc) as the intersection of the steepest tangent (dM/dT) to the M-T curve with the T axis (see supplementary Fig. S8 for more detail). As plotted in the inset of Fig. 3(a), Tc shows an atomic-layer thickness dependence. The Tc for MnSn single layer is ~54 K, and sharply increases to ~225 K for the three-layered MnSn. On the other hand, Tc of the four-layered MnSn is ~235 K, which is only slightly higher than that of three-layered MnSn. The similar behavior has been also reported in the vdW materials of Cr$_2$Ge$_2$Te$_6$[22], Fe$_3$GeTe$_2$[2, 23, 32] and metallic ferromagnetic films[33, 34]. It may imply that samples with different number of layers respond differently to the thermal fluctuations and thus the interlayer magnetic coupling plays a significant role in the ferromagnetic order. A simple model of spin-spin interaction can be taken into account for the rapid decrease of Tc in the few-layer films[33-35]. When the film thickness is below the length of spin-spin correlation, Tc rapidly decreases with decreasing thickness due to the reduced number of pairwise spin-spin interactions[33-

[35]. As a result, the required thermal fluctuations to destroy long range ferromagnetic order would be greatly decreased, and a crossover from 3D to 2D thus appears in the thickness dependent Tc.

The magnetic hysteresis loops (M~H) measurements provide further evidence of the ferromagnetism in monolayer MnSn. Adopting the same method as mentioned in reference[36], the diamagnetic and paramagnetic components are extracted and subtracted (see Fig. S7 in supporting information). As shown in Fig. 3(b), M versus H for the monolayer sample at temperatures below Tc shows a narrow hysteresis loop, which shrinks to disappear above Tc. An in-plane magnetic easy axis for monolayer MnSn is demonstrated by the magnetic anisotropy characterization as displayed in Fig. S9. As the thickness increases, particularly to 4L thick as shown in Fig. S9, there is not a prominent magnetic anisotropy. The ferromagnetic moment is estimated to be ~ 2.3~2.4 $\mu_B$ per unit cell (each Mn atom). This value is also comparable to that measured in another ferromagnetic monolayers of $MnSe_x$[24]. Together with the non-zero remnant magnetization at zero field, it is suggested that an intrinsic ferromagnetism is formed in the MnSn monolayer. Unlike the anti-ferromagnetic interlayer coupling in most 2D ferromagnetic layers, such as $CrI_3$, the saturation magnetization (Ms) of MnSn with odd or even layers exhibit no vanishing net magnetic moment, which indicates the interlayer magnetic coupling is ferromagnetic, as shown in Fig. 3(f). The small coercive force (Hc) of ~450 Oe at 2 K indicates a soft ferromagnetic nature of the monolayer MnSn[37]. As plotted in Fig. 3(e), the Hc shows two–slope dependence on temperature, i.e., the large slope for T < ~10 K and small for > ~10 K. The thicker samples of 3L and 4L show different magnetic behaviors without a sharp enhancement of Hc at low temperature, as displayed in Figs. 3(c)-3(e). This difference between monolayer and 3L or 4L resembles to that observed in magnetization curves, which also indicates the dimensional crossover. Considering the control of film thickness, this enables the capability of tuning the magnetism in MnSn films.

In conclusion, we discovered a new monolayer MnSn epitaxially grown on top of silicon substrate with atomic layer precision, and also demonstrated the ferromagnetism in these films. The magnetic behavior shows clear thickness dependence, which thus provides a strategy to precisely tune the magnetism. This system provides a new platform to study 2D magnetism, and the possibility to integrate 2D ferromagnetism with silicon technology.

**Acknowledgements**

This work was supported by the National Natural Science Foundation of China (Grants No. 11774149, 11790311, 11574133 and 11834006), the National Key R&D Program of China (2016YFA0300404, 2015CB921202 and 2014CB921103). The calculations were carried out using supercomputers at the High Performance Computing Center of Collaborative Innovation Center of Advanced Microstructures, the high-performance supercomputing center of Nanjing University and 'Tianhe-2' at NSCC-Guangzhou.